\documentclass[aps,prl,twocolumn,superscriptaddress,showpacs]{revtex4}
\usepackage{graphicx}
\graphicspath{{${HOME}/fig/publi/muon/}}

\begin{document}


\title{Magnetic density of states at low energy in geometrically 
frustrated systems}



\author {A. Yaouanc}
\affiliation{CEA/DSM/D\'epartement de Recherche
Fondamentale sur la Mati\`ere Condens\'ee, 38054 Grenoble, France}

\author{P. Dalmas de R\'eotier}
\affiliation{CEA/DSM/D\'epartement de Recherche
Fondamentale sur la Mati\`ere Condens\'ee, 38054 Grenoble, France}

\author{V. Glazkov}
\affiliation{CEA/DSM/D\'epartement de Recherche
Fondamentale sur la Mati\`ere Condens\'ee, 38054 Grenoble, France}

\author{C. Marin}
\affiliation{CEA/DSM/D\'epartement de Recherche
Fondamentale sur la Mati\`ere Condens\'ee, 38054 Grenoble, France}

\author{P.~Bonville}
\affiliation{CEA/DSM/D\'epartement de Recherche
sur l'Etat Condens\'ee, les Atomes et les Mol\'ecules, 
91191 Gif sur Yvette, France}

\author{J.A. Hodges}
\affiliation{CEA/DSM/D\'epartement de Recherche
sur l'Etat Condens\'ee, les Atomes et les Mol\'ecules, 
91191 Gif sur Yvette, France}

\author{P.C.M. Gubbens}
\affiliation{Department of Radiation, Radionuclides \& Reactors,
Delft University of Technology, 2629 JB Delft, 
The Netherlands}

\author{S. Sakarya}
\affiliation{Department of Radiation, Radionuclides \& Reactors,
Delft University of Technology, 2629 JB Delft, 
The Netherlands}

\author{C. Baines}
\affiliation{Laboratory for Muon-Spin Spectroscopy, 
Paul Scherrer Institute, 5232 Villigen-PSI, Switzerland}



\date{\today}

\begin{abstract}
Using muon-spin-relaxation measurements we show that the pyrochlore compound 
Gd$_2$Ti$_2$O$_7$, in its magnetically orderered phase below $\sim 1$ K, 
displays persistent spin dynamics down to temperatures as low as
20 mK. 
The characteristics of the induced muon relaxation can be accounted for by 
a scattering process involving two magnetic excitations, with a density of 
states characterized by an upturn at low energy and a
small gap depending linearly on the temperature. 
We propose that such a density of states is a generic feature
of geometrically frustrated magnetic materials.

\end{abstract}


\maketitle

The study of geometrically frustrated materials (GFMs) is a subject at the 
forefront of 
research in condensed matter physics not only because of their own interest 
but also 
because the concept of the frustration of the interactions plays a role for 
understanding
the physics of e.g. ice, cholesteric crystals and metallic 
glasses; see for instance Ref.~\cite{Ramirez01} for a discussion.

Magnetic materials based on lattices with triangular motifs and 
nearest-neighbor antiferromagnetic exchange interaction belong to the
family of GFMs. They are, in the absence of further terms in the 
expression of their energy, believed to remain disordered and fluctuating 
down to zero 
temperature \cite{Villain79}. The absence of magnetic order stems from their 
highly 
degenerate ground state. It is only upon inclusion of perturbations  
such as exchange interactions extending beyond nearest-neighbor magnetic
ions or dipole coupling that magnetic ordering may appear. 
Experiments on Kagom\'e, garnet and pyrochlore 
structure compounds support these general predictions \cite{Ramirez01}.

A fingerprint for the geometrical frustration of a material is the shift
towards low energy of the spectral weight of excitations \cite{Ramirez01}.
The first convincing experimental proof has been obtained in a Kagom\'e-like 
magnetic material, where the low temperature specific heat \cite{Ramirez00}
is found to be dominated by singlet excitations arising from correlated spins,
rather than from individual spins. This enhanced density of spectral weight
at low energy could be linked with the persistence of spin dynamics observed
at low temperature in many systems, as magnetic excitations are continuously
available from zero energy. This behavior has been evidenced in the Kagom\'e 
compounds SrCr$_{9p}$Ga$_{12-9p}$O$_{19}$ with $0.39 \leq p \leq 0.89$ 
\cite{Uemura94,Keren00}, for which a spin-glass transition is detected at 
$T_{\rm g} = 4 p$ K, in Kagom\'e-like systems \cite{Fukaya03,Bono04}, 
in the garnet Gd$_3$Ga$_5$O$_{12}$ \cite{Dunsiger00,Marshall02,Bonville04} 
with $T_{\rm g} = 0.15 $ K and in the pyrochlore compounds Y$_2$Mo$_2$O$_7$ 
($T_{\rm g} = 22 $ K) \cite{Dunsiger96}, 
Tb$_2$Mo$_2$O$_7$ ($T_{\rm g} = 25 $ K) \cite{Dunsiger96}. It was also
observed in the spin liquid Tb$_2$Ti$_2$O$_7$ 
\cite{Gardner99,Gardner03,Keren04}, for which there is no evidence of a 
transition down to 0.07 K \cite{Gardner01}.
The spin dynamics becomes approximately temperature independent below about 
$T_{\rm g}$ for the spin-glass 
systems, except for the two molybdates 
for which it occurs in the temperature range $T/T_{\rm g} <$  0.05, and
for Tb$_2$Ti$_2$O$_7$ which displays a temperature independent relaxation 
below about 1 K.
Appreciable spin dynamics is also found in Yb$_2$Ti$_2$O$_7$ below the 
temperature at which the 
specific heat presents a sharp anomaly \cite{Hodges02a,Yaouanc03,Gardner04b}. 
In compounds where the fluctuations have been studied by the muon spin 
relaxation ($\mu$SR) technique, the muon spin-lattice relaxation function is 
usually found to be a stretched exponential, {\sl i.e.}
$P_Z^{\rm sl}(t) = \exp[-(\lambda_Z t)^\alpha]$ with $\alpha \ne 1$.
 
More unconventional is the spin dynamics recently observed by M\"ossbauer 
spectroscopy, down to 30 mK, in Gd$_2$Sn$_2$O$_7$ 
which shows long-range ordering below 1 K \cite{Bertin02}, and which was 
confirmed by $\mu$SR measurements down to 20 mK \cite{Bonville04a, Dalmas04}. 
This anomalous spin dynamics cannot arise from conventional magnons, since 
their population vanishes at low temperature. In order to further investigate
these zero temperature fluctuations, we performed $\mu$SR measurements in 
the parent compound Gd$_2$Ti$_2$O$_7$, for which single crystals are 
available. This compound undergoes a first magnetic transition at 
$T_{\rm c1} \simeq$ 1~K, followed 
by a second one at $T_{\rm c2} \simeq$ 0.75~K \cite{Ramirez02}. 
We report the results of specific heat and $\mu$SR measurements in this
material, and propose to account for the observed persistent spin
dynamics in GFMs in terms of an unconventional density of states at low 
energy.

A Gd$_2$Ti$_2$O$_7$ polycrystalline rod was prepared by mixing, heating and 
compacting the constituent oxides Gd$_2$O$_3$ and TiO$_2$ of respective 
purity 5N and 4N5. 
A single crystal was then grown by the traveling solvent floating zone
technique using a Crystal System Inc. optical furnace with a velocity
of 8 mm per hour. Oriented platelets were cut from the crystal and 
subsequently annealed under oxygen pressure to ensure optimized physical 
properties. The top panel of Fig. \ref{Gd2X2O7_lambdaZ} shows
the result of specific heat measurements performed using a dynamic adiabatic 
technique. The two phase transitions are observed at $T_{\rm c1}$ = 1.02~K
and $T_{\rm c2}$ = 0.74~K respectively in agreement with recent single 
crystal measurements \cite{Petrenko04}. The peak at $T_{\rm c2}$
is found much sharper than previously reported  
\cite{Ramirez02,Bonville03,Petrenko04}. Its shape suggests that the lower
phase transition is first order. Although we do not have any experimental 
evidence, it is possible that at this lower magnetic transition 
corresponds a structural transition which could be
induced through magneto-elastic coupling. Anyhow, the 
magnetic transitions do not relieve the frustration since in the following 
we do report the observation of persistent
spin dynamics, a fingerprint of geometrical magnetic frustration.
At low temperature, up to $\sim 0.55$~K, the specific heat divided by 
temperature $C_{\rm p}/T$ is proportional to $T$.
Such a behavior was also found for Gd$_2$Sn$_2$O$_7$ 
\cite{Bonville03} and Kagom\'e-like compounds \cite{Ramirez00}, but it 
is not a general rule for GFMs since a $T^2$ dependence is 
reported for Er$_2$Ti$_2$O$_7$ \cite{Champion03}. 
The entropy release reaches $\simeq$ 90 \% of $R\ln 8$ only near 5 K.

\begin{figure}
\includegraphics[scale=0.8]{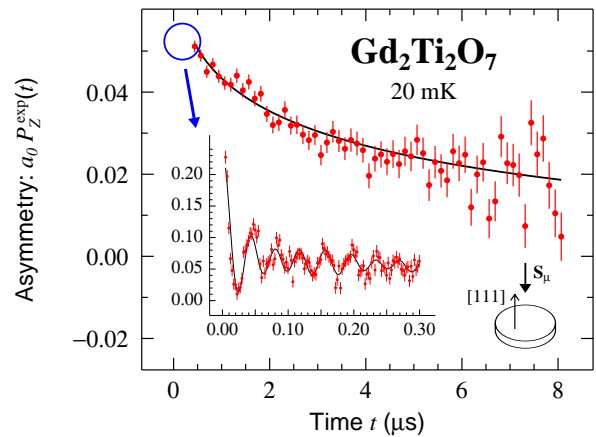}
\caption{(color online)
$\mu$SR spectrum recorded in zero field at 20 mK in single crystals 
of Gd$_2$Ti$_2$O$_7$
with the initial muon beam polarization parallel to the $[111]$ crystal
direction. 
The main frame displays the time spectrum up to 8 $\mu$s, and the insert
magnifies the short time behavior. The latter is 
characterized by wiggles which are damped
out at the earliest time displayed in the main frame.
The wiggles correspond to the spontaneous precession of the
muon spin in two local fields of respective magnitude 
$B_1 = 194\, (1)$  and $B_2 = 149 \, (1)$ mT \cite{note_wiggle}.
As expected, the spin-lattice relaxation channel, most easily seen in the main
frame, accounts for $\sim 1/3$ 
of the spectrum and the wiggles for $\sim \, 2/3$, this being temperature independent
as is also the total asymmetry. The amplitude of the $B_1$ component is 
approximately four times as large as that of $B_2$.
}
\label{Gd2Ti2O7_spectrum}
\end{figure}

\begin{figure}
\includegraphics[scale=0.8]{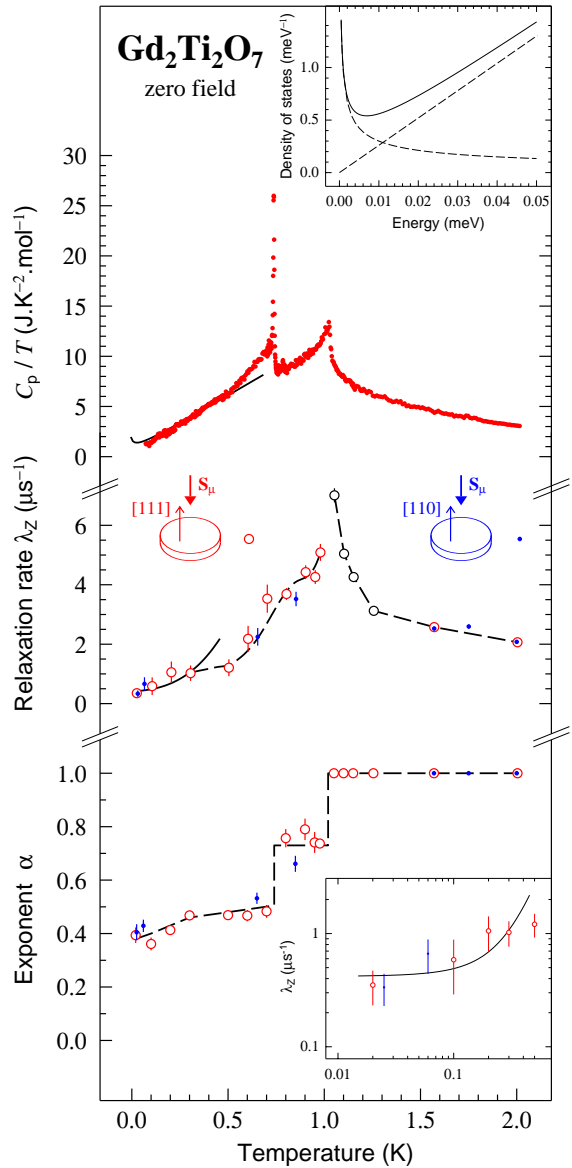}
\caption{(color online) In Gd$_2$Ti$_2$O$_7$, low temperature dependence of 
specific heat over temperature $C_{\rm p}/T$ per mole of Gd (top panel), 
of the muon spin-lattice relaxation rate $\lambda_Z$ (middle panel)
and of the exponent $\alpha$ (bottom panel) of the stretched exponential 
decay.
The muon data were measured in a single crystal, with different 
orientations relative to the muon polarization ${\bf S}_\mu$, as shown, and 
the anisotropy of $\lambda_Z$ was found to be negligible. 
The dashed lines for $\lambda_Z(T)$ and $\alpha(T)$ are guides for the eye. 
The lower insert shows, on a double logarithmic scale, the weak temperature 
dependence of $\lambda_Z(T)$ at low temperature. 
The full lines give the predictions for $C_{\rm p}/T$ and 
$\lambda_Z$ computed with
$b_\mu = $ 0.03 meV$^{-1/2}$, $b_{\rm sh}$
= 26 meV$^{-2}$ per Gd atom and $a$ = 0.1 (see text).
The upper insert shows the density of magnetic excitations $g_m (\epsilon)$
at 50 mK, with its
two components as dashed lines. The gap is: $\Delta \simeq 5\times 
10^{-4}$ meV. The upturn at low energy 
is responsible for the finite value of $\lambda_Z$ as $T \to 0$. 
}
\label{Gd2X2O7_lambdaZ}
\end{figure}

The $\mu$SR measurements (see Ref.~\cite{Dalmas97} for an introduction to 
this technique) were performed at the Low Temperature 
Facility ($\pi$M3 beamline) of the Swiss Muon 
Source (Paul Scherrer Institute, Villigen, Switzerland).
As expected, the $\mu$SR spectra in the paramagnetic phase are well described 
by a single exponential 
relaxation function, {\sl i.e.} they are proportional to $P_Z^{\rm exp}(t) = 
\exp (- \lambda_Z t)$
where $\lambda_Z$ is the spin-lattice relaxation rate.  
$Z$ labels the direction of the initial muon beam polarization. Below 
$T_{\rm c1}$, the observed oscillations, due to spontaneous precession of the
muon spin (see insert of Fig.~\ref{Gd2Ti2O7_spectrum}),
are a signature of the long-range order of the magnetic structure,
consistent with the observation of magnetic Bragg reflections
by neutron diffraction  \cite{Champion01,Stewart04}.
Surprisingly, even at 20 mK \cite{note2}, a stretched exponential decay of the 
spin-lattice relaxation channel is observed, superposed on the damped 
wiggles (Fig.~\ref{Gd2Ti2O7_spectrum}). This implies that excitations of 
the spin system are present,
inducing spin-lattice relaxation of the muon levels. In Fig.~2 are displayed
the thermal variations of the two parameters accounting for this channel,
$\lambda_Z$ and the stretched exponent $\alpha$.

Approaching $T_{\rm c1}$ from above, $\lambda_Z$ increases, reflecting the 
slowing down of the 
paramagnetic fluctuations. As expected, it drops when crossing $T_{\rm c1}$,
but it remains roughly temperature independent below 0.5 K with a value 
$\lesssim$ 1 $\mu$s$^{-1}$. So the persistent spin dynamics known to 
exist in GFMs 
with no long-range magnetic order is also present in a magnetically ordered
compound. Interestingly, $\alpha$ is not far from  1/2 for $T<T_{\rm c2}$ 
and jumps to $\sim 3/4$ for $T_{\rm c2}< T < T_{\rm c1}$. It is expected to 
be equal to 1 for a homogeneous spin system, as found in the paramagnetic 
state. A value $\alpha =1/2$ suggests that relaxation stems from only a 
small fraction $c$ of the spins \cite{McHenry72}, fluctuating with a 
correlation time $\tau_c$ and distributed at random in the
lattice. These spins create a field distribution at the 
muon site which is known to have a squared Lorentzian shape \cite{Walker80}, 
with a width $\Delta_{\rm Lor}$. A recent neutron
diffraction study of Gd$_2$Ti$_2$O$_7$ \cite{Stewart04} has shown that 
$1/4$ of the Gd magnetic moments are only partially ordered in the 
low temperature phase,
and we tentatively attribute the relaxation channel with $\alpha$ = 1/2 to
a fraction of these spins. Our measurements at 0.1 K
with a longitudinal field (not shown) yield $\tau_c = 0.7 \, (2) $ ns and the
relationship $\lambda_Z$ = $4\Delta_{\rm Lor}^2\tau_c$ yields a width
$\Delta_{\rm Lor} = 18 \, (5)$ mT. 
According to Uemura {\it et al.} \cite{Uemura85}, 
$\Delta_{\rm Lor} = \sqrt{\pi/2} c \Delta_{\rm max}$, where $\Delta_{\rm max}$
is the field width if all the Gd moments were contributing to the muon 
relaxation. Using the 
scaling between $\Delta_{\rm max}$ and the rare earth moment which was 
successfully used for Tb$_2$Ti$_2$O$_7$ \cite{Gardner99} and Yb$_2$Ti$_2$O$_7$
\cite{Yaouanc03} we deduce $c \lesssim$ 10\%{}. We note that a 
significant distribution in the
spin fluctuation times has been observed, by the neutron spin echo
technique, in Gd$_2$Ti$_2$O$_7$ \cite{Gardner04a} and in Tb$_2$Ti$_2$O$_7$ 
\cite{Gardner03}. This backs our observation that only a fraction 
of the Gd$^{3+}$ ions
relax the muon spin, whereas the remaindering part does not because it
is characterized by fluctuation times either too short or too long.

Our specific heat and $\mu$SR measurements show therefore that, in the long
range order phase of Gd$_2$Ti$_2$O$_7$, magnetic excitations with a 
non-vanishing density at low energy are present. These excitations lead
to a $T^2$ behaviour for the specific heat and to a muon spin lattice
relaxation rate which is quasi-independent of temperature. 
In fact below 0.5 K a fit of $\lambda_Z(T)$ to a power law gives
$\lambda_Z(T) \propto T^\beta$ with $\beta \simeq 1/3$.
This is a negligible temperature dependence relative to power laws
observed for usual ferromagnets ($\beta = 2$) and conventional 
antiferromagnets ($\beta = 5$) \cite{Dalmas04}. In the following, we aim at
determining the density of magnetic excitations responsible for the observed
thermal behaviours of $C_p$ and $\lambda_Z$. 

First of all, assuming these excitations to
obey Bose-Einstein statistics like conventional magnons, it is easy to derive
that, if their density of states per volume unit $g_m(\epsilon)$ is 
proportional to $\epsilon^q$, then
$C_p \propto T^{q+1}$. Hence $g_m(\epsilon) \propto \epsilon$ accounts for
the $T^2$ behaviour of $C_p$. Second, for the muon relaxation rate, if one
considers a direct process with a single excitation, then energy conservation
with $\epsilon=\epsilon_\mu \simeq 0$ ($\epsilon_\mu = \hbar 
\omega_\mu$, $\omega_\mu$ being the muon angular frequency) leads to
$\lambda_Z \propto T$, which is in disagreement with the experimental 
data for Gd$_2$Ti$_2$O$_7$ and other frustrated systems mentioned in the
introduction. The 
relaxation process to consider next involves a {\sl two-excitation scattering 
(Raman process)}. Within the harmonic approximation, energy
conservation implies that only the component of the spin-spin correlation
tensor parallel to the magnetization,  $\Lambda^{\parallel}({\bf q}, 
\omega_\mu=0)$, is probed \cite{Dalmas95}. Thus,  
$\lambda_Z \propto \int \mathcal{C}({\bf q}) \Lambda^{\parallel}({\bf q}, 
\omega_\mu=0) \, {\rm d}^3{\bf q}$, where $\omega_\mu$ is set to zero because
it is vanishingly small, $\mathcal{C}({\bf q})$ accounts for the 
interaction of the muon spin with the lattice spins and the integral extends 
over the Brillouin zone. For antiferromagnets with two collinear sublattices, 
the muon relaxation rate due to a Raman magnon process has been derived
assuming for simplicity no orientation dependence for 
$\epsilon(q)$ and $\mathcal{C}(q)$:
\begin{eqnarray}
\lambda_ Z & = & {8 (2\pi)^3 \mathcal{D}\hbar  \over 15}
                 { (B_e+B_a)^2 \over (B_e+B_a)^2 -B_e^2 }\cr 
&\times & \int_{\Delta}^{\infty}  n \left (\epsilon/k_{\rm B}T  \right ) 
\left [n \left (\epsilon/k_{\rm B}T  \right ) +1  \right ]  
g_m^2(\epsilon) {\rm d} \epsilon,
\label{eq_lambda_hm}
\end{eqnarray}
where ${\mathcal D}$ = $(\mu_0/4\pi)^2 \gamma_\mu^2 g^2 \mu_B^2$, 
$n (\epsilon /k_{\rm B}T )$ is the Bose-Einstein 
occupation factor, and ${\Delta}$ the energy gap of the excitations at zero
energy. $B_e$ and $B_a$ are respectively the exchange and anisotropy fields;
a mean-field estimate for $B_e$ is 10~T and, with reference to another
Gd compound \cite{Yaouanc96}, we take $B_a$ = 0.2~T.
Equation \ref{eq_lambda_hm}
contains two population factors standing for the 
creation and annihilation of an excitation, a density of states being 
associated with each of them.
The expression given by Eq.~\ref{eq_lambda_hm} assumes a muon site of
high symmetry, e.g. the octahedral or tetrahedral interstitial sites of
a face centered cubic lattice. For a muon site of lower symmetry, as expected
in the case of the pyrochlores, Eq.~\ref{eq_lambda_hm} is essentially modified
by a multiplicative factor $\eta$ which depends on the actual site and is in
the range between 1 and $\sim$ 10. We took $\eta$ = 7 in the calculation 
below.

As a first step, let us assume $\lambda_ Z$ to be temperature independent. 
The density of states must be $g_m(\epsilon) = b_\mu \epsilon^{-1/2}$ \cite{note1}. It must also 
be assumed that ${\Delta} = a k_{\rm B} T$, i.e. the gap at zero energy is
proportional to temperature. If $a$ is of order 1 or lower, it can be shown 
that $\lambda_Z\propto b_\mu^2/a^2$ and that $g_m(\epsilon)$ is essentially
probed for $\Delta \leq \epsilon \lesssim 3 \Delta$. Therefore the 
$\epsilon^{-1/2}$ dependence for $g_m(\epsilon)$  
holds only in a very restricted energy interval. As stated above, the $T^2$
dependence of $C_{\rm p}$ implies that $g_m(\epsilon)$ is linear with energy,
and this must prevail for $\epsilon \gtrsim 3\Delta$. 
Combining the two regimes, we obtain: 
$g_m(\epsilon) = b_\mu \epsilon^{-1/2} + b_{\rm sh} \epsilon$. 
The numerical calculations show that $a \leq 0.1 $ is compatible with the 
$C_{\rm p}/T$ and $\lambda_Z$ data. 
The predictions of the model with $a$ = 0.1 are shown in 
Fig.~\ref{Gd2X2O7_lambdaZ}. As expected the model predicts $\lambda_Z$ to be
independent of $T$ at low temperature. Note that a further refinement of the 
model, namely a slight change in the value 
of the exponent in the former term of
$g_m(\epsilon)$ ($-0.5\to -0.4$), allows to improve the fit of $\lambda_Z(T)$
at low $T$ for the specific case of Gd$_2$Ti$_2$O$_7$. This change 
affects the
details in the shape of $g_m(\epsilon)$, but not its main features:
the upturn at small energy and the existence of a gap proportionnal to the
temperature.

The calculated $C_{\rm p}/T$ shown in Fig.~\ref{Gd2X2O7_lambdaZ} 
is $T$-linear as
expected, but shows an upturn below about 40 mK, caused by the 
$\epsilon^{-1/2}$ dependence of the density of states. This signature
could be searched for in very low temperature specific heat measurements 
performed on a sample enriched with Gd isotopes having zero nuclear moment, 
thus showing no nuclear Schottky anomaly.
The computed curve for $\lambda_Z$ above $\sim 0.4$ K  
overestimates the measured values. This could indicate that a fraction 
of the density of states, in the region where the dependence  
$g_m(\epsilon) \propto \epsilon$ prevails, arises from singlet-like states
\cite{Ramirez00} which do not contribute to the relaxation.

In summary we have shown that, in GFMs, a non-vanishing muon spin-lattice
relaxation at low $T$, with a weak temperature dependence, if any,
is the signature of a low energy upturn in the
density of magnetic states. This density is characterized by a gap 
varying linearly with temperature, leading to an accumulation of 
states at low energy (see the upper insert of Fig.~2).
This is a rare feature, which is also observed 
in BCS superconductors. We propose that the gap 
is due to the dipole interaction between magnetic moments,
which breaks rotational invariance. 
However, its increase with $T$ is unexpected. It is usually
temperature independent or, for superconductors, decreases as temperature is 
increased \cite{Anderson84}. A linear thermal increase strongly suggests that  
the thermal energy exceeds the energy involved in the excitation 
scattering process. Indeed, an
analytical classical calculation performed for a triangular planar model with 
nearest-neighbor antiferromagnetic interactions and dipole interactions 
predicts a gap proportional to $T$  with $a = 0.2$ for one of the phases 
\cite{Rastelli03}.

As pointed out at the beginning of this Letter, persistent and weakly
temperature spin dynamics has been observed for a large number of GFM, included magnetically 
ordered powder samples of Gd$_2$Ti$_2$O$_7$ \cite{Yaouanc03}, Gd$_2$Sn$_2$O$_7$ 
\cite{Bonville04, Dalmas04} and Er$_2$Ti$_2$O$_7$ \cite{Lago05}. A combined 
analysis of $C_p(T)$ and $\lambda_Z(T)$ is always possible, resulting in $g_m$ with a gap 
linear in temperature.

The possibility for a GFM to order magnetically at finite temperature,
although it is predicted to remain disordered down to zero temperature, 
was discovered theoretically a long time ago \cite{Villain80}, as ``order 
from thermal disorder''. The gap we infer here is a consequence of the same 
mechanism. It appears if rotational invariance is broken.
This naturally occurs because of the presence of the dipole interaction.

We thank B. Canals, G. Jackeli, S.V. Maleyev for 
enlightning discussions. Part of this work was performed at the
S$\mu$S, Paul Scherrer Institute, Villigen, Switzerland.

\bibliography{density_letter}

\begin{thebibliography}{42}
\bibitem[{not({\natexlab{a}})}]{note2}
\bibinfo{note}{We have no direct proof of the sample thermalization down to 20
  mK. However, using the same dilution refrigerator we have always observed by
  $\mu$SR signatures of phase transitions where they were expected, e.g. at 240
  mK in Ref. \cite{Hodges02a} and 54 mK in Ref. \cite{dalmas03}. Using single
  crystals as in the present case, one expects sample thermalization to be at
  least as good as for the powder samples studied in both references.}
\bibitem[{not({\natexlab{b}})}]{note_wiggle}
\bibinfo{note}{The details of the function used to fit the precessing part of
  the experimental spectra will be published elsewhere}.
\bibitem[{not({\natexlab{c}})}]{note1}
\bibinfo{note}{It is possible to get $\lambda_Z$ temperature independent using
  the conventional dispersion relation for antiferromagnets $\epsilon = c_{\rm
  AF} \, q$ with $c_{\rm AF} \propto T^{5/6}$. However, this leads to a
  specific heat proportional to $T^{1/2}$, in disagreement with experiment. The
  proportionality factor leads also to much higher values for the low
  temperature specific heat than actually measured.}
\end{thebibliography}

\end{document}